\documentclass[%
 aps,
pra,%
 amsmath,amssymb,
 reprint,%
]{revtex4-1}

\usepackage{graphicx}% Include figure files
\usepackage{dcolumn}% Align table columns on decimal point
\usepackage{bm}% bold math
\begin{document}

\title{X-ray Multi-modal Intrinsic-Speckle-Tracking}% Force line breaks with \\

\author{Konstantin M. Pavlov}
\email{konstantin.pavlov@canterbury.ac.nz}
\affiliation{School of Physical and Chemical Sciences, University of Canterbury, Christchurch, New Zealand}
\affiliation{School of Physics and Astronomy, Monash University, Victoria 3800, Australia}
\affiliation{School of Science and Technology, University of New England, NSW 2351, Australia}

\author{David M. Paganin}
\affiliation{School of Physics and Astronomy, Monash University, Victoria 3800, Australia}

\author{Heyang (Thomas) Li}
\affiliation{School of Mathematics and Statistics, University of Canterbury, Christchurch, New Zealand}
\affiliation{School of Physics and Astronomy, Monash University, Victoria 3800, Australia}

\author{Sebastien Berujon}
\affiliation{Instituto COPPEAD de Administra{\c{c}}\~ao, Universidade Federal do Rio de Janeiro, Rio de Janeiro, Brazil}
%\affiliation{European Synchrotron Radiation Facility, 38043 Grenoble, France}

\author{H\'{e}l\`{e}ne Roug\'{e}-Labriet}
\affiliation{Novitom, 3 av doyen Louis Weil 38000 Grenoble, France}
\affiliation{Inserm UA7 STROBE, Universit\'{e} Grenoble Alpes, 38000 Grenoble, France}

\author{Emmanuel Brun}
\affiliation{Inserm UA7 STROBE, Universit\'{e} Grenoble Alpes, 38000 Grenoble, France}

\date{\today}% It is always \today, today,
             %  but any date may be explicitly specified

\begin{abstract}
We develop X-ray Multi-modal Intrinsic-Speckle-Tracking (MIST), a form of X-ray speckle-tracking that is able to recover both the position-dependent phase shift and the position-dependent small-angle X-ray scattering (SAXS) signal of a phase object.  MIST is based on combining a Fokker--Planck description of paraxial X-ray optics, with an optical-flow formalism for X-ray speckle-tracking.  Only two images need to be taken in the presence of the sample, corresponding to two different transverse positions of the speckle-generating membrane, in order to recover both the refractive and local-SAXS properties of the sample.  Like the optical-flow X-ray phase-retrieval method which it generalises, the MIST method implicitly rather than explicitly tracks both the transverse motion and the diffusion of speckles that is induced by the presence of a sample.  Application to X-ray synchrotron data shows the method to be efficient, rapid and stable. 
\end{abstract}
% KMP 24_08_20: replaced "refractive index decrement" by "phase shift" in the abstract above
\maketitle

\section{Introduction}

In two recent papers \cite{MorganPaganin2019,PaganinMorgan2019} a Fokker--Planck \cite{Risken1989} formalism was developed for paraxial X-ray optics.  The essence of this formalism is to use a two-current continuity equation, to describe paraxial X-ray energy flow downstream of illuminated samples containing both spatially-resolved phase--amplitude fluctuations and spatially-unresolved random micro-structure.  The unresolved sample micro-structure bifurcates the X-ray energy transport, both within and downstream of the illuminated sample, into coherent and diffuse channels. The coherent energy-flow channel, downstream of the illuminated sample, is associated with the spatially resolved attenuation and refractive properties of the sample.  The diffuse energy-flow channel is associated with ``local roughness'' \cite{GureyevPMB2020} modelled by the spatially unresolved random micro-structure within the sample.   

The paraxial-optics Fokker--Planck equation \cite{MorganPaganin2019,PaganinMorgan2019} is an elliptic second-order partial differential equation that may be viewed as a generalised form of Teague's transport-of-intensity equation \cite{teague1983} for coherent paraxial optics.  This generalisation simultaneously incorporates the additional effects of local incoherent scatter (small-angle X-ray scattering, SAXS \cite{GlatterKratky1982}), source-size blurring and detector-induced blurring \cite{MorganPaganin2019,PaganinMorgan2019}. Such a Fokker--Planck approach to paraxial X-ray optics in its final formulation is somewhat similar to statistical dynamical diffraction theory (SDDT), developed in the 1980s and 1990s by Kato \cite{Kato1,Kato2} and further developed by others \cite{PavlovPunegov1998,PavlovPunegov2000,Nesterets2000}, to describe dynamical and kinematical diffraction by deformed crystals having chaotically distributed defects. Later a similar statistical approach was applied by \citeauthor{Nesterets2008} \cite{Nesterets2008} in the context of phase-contrast X-ray imaging (PCI) of non-crystalline objects.  Further parallels include diffuse X-ray scattering from crystals \cite{WarrenBook}, the frozen phonon model of electron diffraction \cite{KirklandBook}, optical scattering from rough surfaces \cite{VoronovichBook} and radiative transport in turbid media \cite{Khelashvili2005,OlbrantFrank2010}. 

The smallness (or high concentration) of either crystal defects (in SDDT) or object features (in PCI) (here ``smallness'' is in comparison with the resolution of the detection system) requires one to apply a statistical approach via averaging over a statistical ensemble to describe scattering by some ``unresolvable'' elements of such systems. Scattering by such ``unresolvable'' features transfers the propagating energy (intensity) from the coherent channel into the diffuse channel (see also Chap~7.4 in the book by \citeauthor{IshimaruBook1978} \cite{IshimaruBook1978}). In the context of PCI, the effect of the diffuse component is usually described in terms of broadening caused by SAXS \cite{Pagot2003,Rigon2003,Wernick2003,Khelashvili2005,Pfeiffer2008,Kitchen2010,BerujonWangSawhney2012,Endrizzi2014}. However, a division into coherent and diffuse components was also intrinsically used (see e.g., Eq.~(1) in \citeauthor{Oltulu2003} \cite{Oltulu2003}). This transfer of X-ray energy from coherent intensity into diffuse intensity may be comparable with photoelectric absorption loss of this coherent component if the concentration of such defects (in SDDT) \cite{Pietsch2004Book} or features (in PCI) \cite{Nesterets2008}  is high. The diffuse component of intensity can be further re-scattered if the object is thick enough. However, such typically small dynamical effects are neglected for diffuse intensity \cite{Bushuev1989}.

A separate but related thread of development is X-ray speckle-tracking \cite{berujon2012,morgan2012}.  In this X-ray imaging method, speckles produced by a spatially random screen are recorded in the presence of a sample.  Comparison of these speckles to those recorded in the absence of the sample, for one or more mask positions, then allows the refractive, attenuating and local-SAXS properties of the sample to be inferred.  See the recent review by \citeauthor{zdora2018} \cite{zdora2018} together with references cited therein, as well as precedent work in a visible-light context \cite{Massig1,Massig2,Perciante}.  Note also the evident similarities to X-ray Hartmann--Shack sensors \cite{MayoSexton2004} and single-grid phase-measurement methods \cite{Wen2010,Morgan2011}, both of which use specially-fabricated rather than random masks.  

Multi-modal recovery of phase, intensity and SAXS is enjoying much attention in an X-ray speckle-tracking context, e.g.~using the ``X-ray Speckle-Vector Tracking'' (XSVT) formalism \cite{Berujon2015c}, and the formalism of ``Unified Modulated Pattern Analysis'' (UMPA) \cite{Zdora2017}.  XSVT and UMPA both employ multiple images for multi-modal recovery, although single-image multi-modal X-ray speckle tracking is also possible, using correlation-based approaches \cite{Morgan2013,Zanette2014}.  There is an evident trade-off here, as, broadly speaking, measuring more images (e.g.~in XSVT and UMPA) gives the benefit of improved spatial resolution at the cost of increased dose to the sample, relative to single-shot approaches.  Another formalism requiring only a single speckle image to be taken in the presence of the sample, termed ``Optical Flow'' (OF) \cite{PaganinLabrietBrunBerujon2018}, has very recently been developed for X-ray speckle tracking, however this last-mentioned approach has not been applied to multi-modal analyses \footnote{``Optical flow'' has a well-established usage in the context of image processing, which is similar to that in the present paper, but does not use inbuilt local energy conservation.  Indeed, our key assumption of local energy conservation---as described by the continuity equation---will not be valid in the more general context of image processing.  See e.g.~\citet{OF_old1} and \citet{OF_old2}, together with references therein.}.  As mentioned earlier, XSVT and UMPA are multi-modal (as are the single-shot methods published in e.g.~Refs.~\cite{Morgan2013,Zanette2014}); the present paper makes OF multi-modal by augmenting it from a method able to recover wave-field phase, into a method able to recover both wave-field phase and local-SAXS properties. 

OF is here made multi-modal via a Fokker--Planck-type generalisation that incorporates local SAXS, not unlike passing from non-statistical to statistical diffraction theory. An attractive feature of the OF formalism is that it implicitly rather than explicitly tracks speckles, making it computationally much more rapid than more-general methods that rely on correlation analyses and/or error-metric minimisation.  Another attractive feature is the complementary approach that the Fokker--Planck extension to OF takes to speckle tracking, being based on a non-iterative closed-form solution of a specified partial differential equation, as distinct from approaches that are based on iteratively minimising an error-metric integral.

We close this introduction with a brief outline of the remainder of the paper.  In Sec.~II we outline the theory underpinning our technique, which we term ``X-ray Multi-modal Intrinsic-Speckle-Tracking (MIST)''.  We then apply this formalism, in Sec.~III, to experimental hard X-ray data obtained at the European Synchrotron Radiation Facility (ESRF).  We briefly discuss some of the broader implications of our results in Sec.~IV, and offer some concluding remarks in Sec.~V.

\section{Theory}

\begin{figure}
\includegraphics[width=0.46\textwidth]{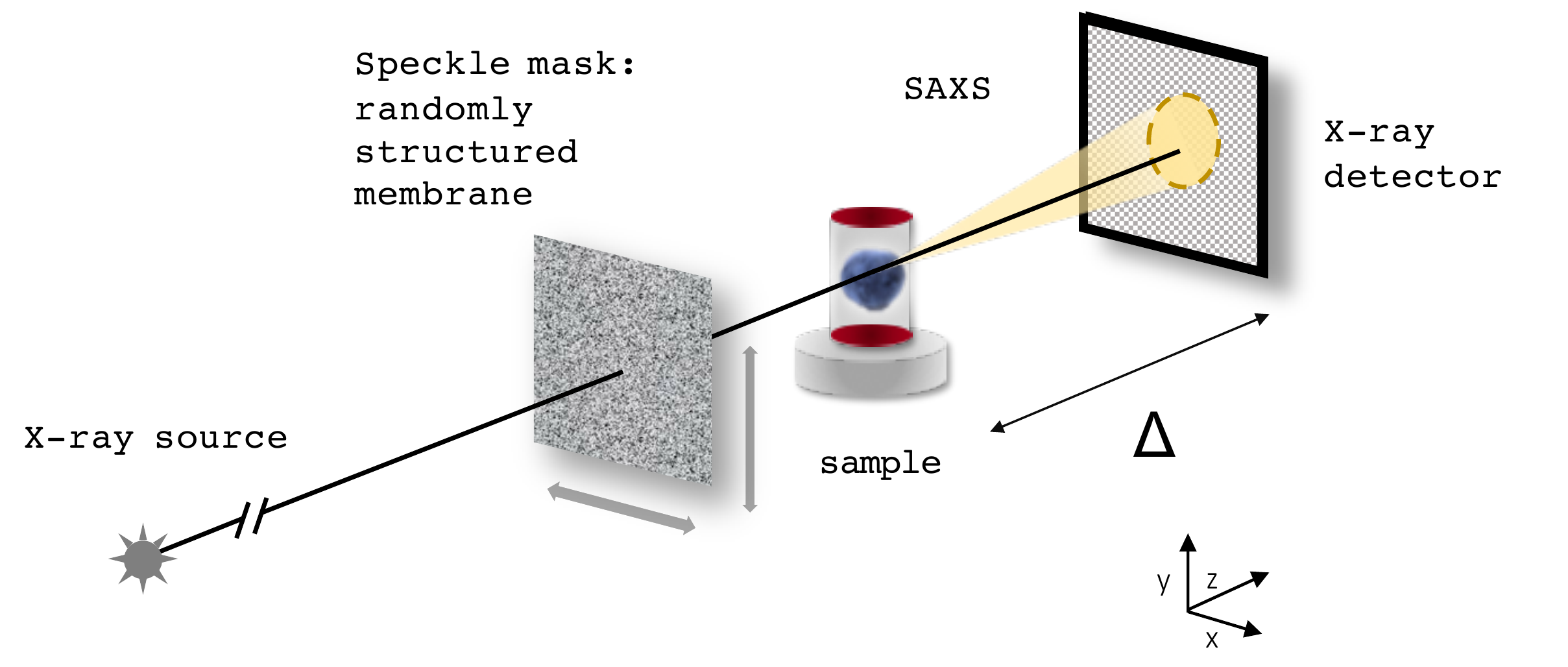}
\caption{Experimental setup for X-ray Multi-modal Instrinsic-Speckle-Tracking.\label{fig:ExperimentalSchematic}}
\end{figure}

Assume that a pure-phase-object sample is placed in a spatially-well-resolved reference speckle field, such as that sketched in Fig.~\ref{fig:ExperimentalSchematic}. The reference speckle field may be created by passing an X-ray beam through a spatially random membrane.  The registered speckle images obey the following Fokker--Planck \cite{Risken1989} generalisation of the OF formalism for speckle-tracking \cite{PaganinLabrietBrunBerujon2018}.  This gives Eq.~(55) in the theoretical paper by \citeauthor{PaganinMorgan2019} \cite{PaganinMorgan2019}, which was proposed but neither solved nor applied in that previous publication, and which which forms the Fokker--Planck-type starting-point for the present paper:
\begin{eqnarray}
  \nonumber I_{\textrm R}(x,y)-I_{\textrm S}(x,y)=\frac{\Delta}{k} \nabla_{\perp}\cdot [I_{\textrm R}(x,y) \nabla_{\perp}\phi(x,y)] \\ -\Delta\nabla_{\perp}^2[D_{\textrm{eff}}(x,y;\Delta)I_{\textrm R}(x,y)].  \quad \label{eq:1} 
\end{eqnarray}
Here, $I_{\textrm R}(x,y)$ is a reference speckle image obtained in the absence of a sample, $I_{\textrm S}(x,y)$ is the corresponding speckle image obtained in the presence of a sample that is by assumption a pure-phase object, $(x,y)$ denote transverse coordinates in planes perpendicular to the optical axis $z$,  $\Delta$ is the sample-to-detector distance, $k$ is the X-ray wave number, $\phi$ is the phase shift caused by the sample, $\nabla_{\perp}\equiv(\partial/\partial x,\partial/\partial y)$ is the transverse gradient and $D_{\textrm{eff}}$ is the effective diffusion coefficient describing local sample-induced SAXS \cite{MorganPaganin2019,PaganinMorgan2019}.  %This diffusion coefficient, which is also termed a ``dark-field'' signal in much of the X-ray and neutron literature \cite{Pfeiffer2008}, is assumed to be a slowly-varying function (i.e., we can neglect its derivatives, which are small). The first term on the right side of Eq.~(\ref{eq:1}) corresponds to the coherently scattered intensity, while the second describes diffuse scattering (local SAXS) that is due to unresolved micro-structure in the sample.

The first term, on the right side of Eq.~(\ref{eq:1}), represents a coherent flow that deforms $I_{\textrm R}(x,y)$ into $I_{\textrm S}(x,y)$, with this deformation being induced by the refractive properties of the sample.  This coherent flow has two distinct components \cite{PaganinLabrietBrunBerujon2018}, associated with the two terms in the expansion:
\begin{eqnarray}
\label{eq:ExpandDivIGradPhi}
\nabla_{\perp}\cdot [I_{\textrm R}(x,y) \nabla_{\perp}\phi(x,y)] \quad\quad\quad\quad\quad\quad\quad\quad\quad\quad\quad  \\ \nonumber = I_{\textrm R}(x,y)\nabla_{\perp}^2 \phi(x,y)  + \nabla_{\perp}I_{\textrm R}(x,y)\cdot\nabla_{\perp}\phi(x,y).
\end{eqnarray} 
Here, the term  $(\Delta/k) I_{\textrm R}(x,y)\nabla_{\perp}^2 \phi(x,y)$ describes a lensing contribution in which speckles may be locally magnified or demagnified on account of the curvature $\nabla_{\perp}^2 \phi(x,y)$ of the object-induced X-ray phase shift.  This term will also describe the propagation-based phase contrast (Laplacian of the phase) induced by edges in the sample \cite{wilkins1996}, which will be `imprinted' on top of the speckle (see e.g.~Ref.~\cite{Groenendijk2020}).  Conversely, the term $(\Delta/k) \nabla_{\perp}I_{\textrm R}(x,y)\cdot\nabla_{\perp}\phi(x,y)$ describes the transverse motion of the speckles that is induced by the sample.  Turning to the diffusive channel of the energy flow, we have the second term on the right-hand side of Eq.~(\ref{eq:1}), which describes the diffusion (visibility reduction) of the speckles that is induced by the unresolved micro-structure within the sample.  This sample-induced diffusion is quantified by the position-dependent effective diffusion coefficient $D_{\textrm{eff}}(x,y;\Delta)$. Thus, Eq.~(\ref{eq:1}) is a partial differential equation embodying three separate physical mechanisms for the sample-induced deformation (``flow'') that maps speckles of the reference image $I_{\textrm R}(x,y)$ to corresponding speckles $I_{\textrm S}(x,y)$ measured in the presence of the sample: (i) lensing of speckles; (ii) transverse motion of speckles; (iii) diffusion of speckles. 

The Laplacian operator, applied to the second term on the right side of Eq.~(\ref{eq:1}), yields the following three components for the diffusive component of the sample-induced speckle deformation:
\begin{eqnarray}
\nonumber  \nabla_{\perp}^2[D_{\textrm{eff}}(x,y;\Delta) && \!\!\!\!\!\!\! I_{\textrm R}(x,y)]   = D_{\textrm{eff}}(x,y;\Delta)   
 \nabla_{\perp}^2 
 I_{\textrm R}(x,y)  \\ \nonumber &+& I_{\textrm R}(x,y)
 \nabla_{\perp}^2 
 D_{\textrm{eff}}(x,y;\Delta)   
 \\  &+& 2 \nabla_{\perp}  D_{\textrm{eff}}(x,y;\Delta)  \cdot \nabla_{\perp} I_{\textrm R}(x,y). \label{eq:2} 
\end{eqnarray}
We can neglect the second and third terms on the right-hand side of Eq.~(\ref{eq:2}) on account of the assumption that the position-dependent effective diffusion coefficient  $D_{\textrm{eff}}(x,y;\Delta)$ is a slowly-varying function of transverse coordinates. We can therefore simplify Eq.~(\ref{eq:1}) as follows:
\begin{eqnarray}
  \nonumber I_{\textrm R}(x,y)-I_{\textrm S}(x,y)=\frac{\Delta}{k} I_{\textrm R}(x,y) \nabla_{\perp}^2\phi(x,y) \\ -\Delta D_{\textrm{eff}}(x,y;\Delta) \nabla_{\perp}^2I_{\textrm R}(x,y),  \quad \label{eq:3} 
\end{eqnarray}
where we have also used the approximation previously employed in \citeauthor{PavlovPhysRevAppl2020} \cite{PavlovPhysRevAppl2020}, namely 
\begin{eqnarray}
\nabla_{\perp} I_{\textrm R}(x,y) \cdot \nabla_{\perp}\phi(x,y)\approx 0.  
\end{eqnarray}

We now explain this last-mentioned approximation in a little more detail.  The intensity $I_{\textrm R}(x,y)$ of the reference speckle image, acquired in the absence of a sample, is produced by a spatially random mask. Therefore, the gradient of such an intensity field will be a vector field that is rapidly changing in both direction and magnitude, as a function of transverse coordinates. Thus, the scalar product of such a random vector field with a more slowly changing gradient of the phase can
be neglected. 
% KMP added (24_08_20) an additional explanation below:
The same approximation can also be explained in geometric terms, by calculating the expectation of the dot product of a pair of two-dimensional vector fields: (i) a random vector field ${\bf{r}}(x,y)$ and (ii) a slowly varying vector field ${\bf{s}}(x,y)$. The expectation (averaged over position coordinates) of such a scalar product ${\bf{r}}(x,y) \cdot {\bf{s}}(x,y)$ is a sum of the expectations of the appropriate products of the projections of these two vectors. The expectation of a product of projections of two vectors is equal to the product of the expectations of each of these projections, as the vector fields ${\bf{r}}(x,y)$ and ${\bf{s}}(x,y)$ are statistically independent. Hence we have a product of zero, which is an expectation of the projection of a random vector field, and the expectation of the projection of the slowly varying vector field. 

So far in this section, we have considered the ``forward problem'' of how the the sample deforms the reference speckle image $I_{\textrm R}(x,y)$ into the speckle image $I_{\textrm S}(x,y)$ that is measured when the sample is present.  This physical model for the forward problem enables us to next consider the associated ``inverse problem'' \cite{Sabatier2000} of inverting the measured intensity data so as to infer information regarding the sample.
To this end, observe that Eq.~(\ref{eq:3}) contains two unknown functions, namely $\nabla_{\perp}^2\phi(x,y)$ and $D_{\textrm{eff}}(x,y;\Delta)$, which can
be recovered using measured intensity data corresponding to two different transverse positions (``\#1'' and ``\#2'') of the mask. Thus we can write a system of simultaneous equations for mask positions \#1 and \#2, based on Eq.~(\ref{eq:3}):
\begin{eqnarray}\label{eq:4} 
  \begin{cases}
              I_{\textrm R_1}(x,y)-I_{\textrm S_1}(x,y)=\frac{\Delta}{k} I_{\textrm R_1}(x,y) \nabla_{\perp}^2\phi(x,y) \\ \quad\quad\quad\quad\quad\quad\quad\quad\quad\quad -\Delta D_{\textrm{eff}}(x,y;\Delta) \nabla_{\perp}^2I_{\textrm R_1}(x,y),\\
               I_{\textrm R_2}(x,y)-I_{\textrm S_2}(x,y)=\frac{\Delta}{k} I_{\textrm R_2}(x,y) \nabla_{\perp}^2\phi(x,y) \\ \quad\quad\quad\quad\quad\quad\quad\quad\quad\quad -\Delta D_{\textrm{eff}}(x,y;\Delta) \nabla_{\perp}^2I_{\textrm R_2}(x,y).  
            \end{cases}
\end{eqnarray}
%
% KMP added "The above system of simultaneous linear equations can be solved algebraically , to obtain both the functions..." in the text below 24_08_20
Here, $I_{\textrm R_{1,2}}(x,y)$ denotes the reference speckle images corresponding to random masks in positions \#1 and \#2, with $I_{\textrm S_{1,2}}(x,y)$ similarly defined. The above system of simultaneous linear equations can be solved algebraically, to obtain both the functions $\nabla_{\perp}^2\phi(x,y)$ and $D_{\textrm{eff}}(x,y;\Delta)$, with the latter quantity being given by:
\begin{eqnarray}\label{eq:5}
D_{\textrm{eff}}(x,y;\Delta) \quad\quad\quad\quad\quad\quad\quad\quad\quad\quad\quad\quad\quad\quad\quad\quad\quad\quad \\ \nonumber = \frac{1}{\Delta} \frac{I_{\textrm S_1}(x,y)I_{\textrm R_2}(x,y)-I_{\textrm S_2}(x,y)I_{\textrm R_1}(x,y)}{I_{\textrm R_2}(x,y)\nabla_{\perp}^2I_{\textrm R_1}(x,y)-I_{\textrm R_1}(x,y)\nabla_{\perp}^2I_{\textrm R_2}(x,y)}.  
\end{eqnarray}
% KMP removed the Laplacian of the phase
% \nabla_{\perp}^2\phi(x,y) &= \frac{k}{\Delta}\frac{[I_{\textrm R_1}(x,y)-I_{\textrm S_1}(x,y)]\nabla_{\perp}^2I_{\textrm R_2}(x,y)-[I_{\textrm R_2}(x,y)-I_{\textrm S_2}(x,y)]\nabla_{\perp}^2I_{\textrm R_1}(x,y)}{I_{\textrm R_1}(x,y) \nabla_{\perp}^2 I_{\textrm R_2}(x,y)-I_{\textrm R_2}(x,y)\nabla_{\perp}^2I_{\textrm R_1}(x,y)}, \\

As $I_{\textrm R_1}(x,y)$ and $I_{\textrm R_2}(x,y)$ are the intensities of a reference speckle image with the random mask in two different transverse spatial positions, it is unlikely that the denominator in Eq.~(\ref{eq:5}) will be close to zero. Therefore, the solution given in Eq.~(\ref{eq:5}) is well defined. 
% KMP rephrased this sentence
% Using boundary conditions for the phase shift, namely that the phase shift is zero outside the sample, one can reconstruct the phase shift from its Laplacian obtained in Eq.~(\ref{eq:5}). 
One can easily recover the Laplacian of the phase shift using Eq.~(\ref{eq:4}). Then the phase shift can be obtained by integrating the resulting expression for the phase Laplacian, using a fast Fourier transform approach (see e.g.~\citeauthor{GureyevNugent1997} \cite{GureyevNugent1997}) to yield $\phi(x,y)$ up an arbitrary additive constant. A similar approach to the recovery of the phase shift $\phi(x,y)$ was used in our previously published OF-based intrinsic-speckle-tracking study \cite{PaganinLabrietBrunBerujon2018}. However, in the present paper we are instead focused on reconstruction of the effective diffusion coefficient $D_{\textrm{eff}}(x,y;\Delta)$ using Eq.~(\ref{eq:1}), since this extends intrinsic speckle-tracking from its present uni-modal capability \cite{PaganinLabrietBrunBerujon2018}, for which phase recovery has already been demonstrated, to a multi-modal technique that is also able to measure $D_{\textrm{eff}}(x,y;\Delta)$. 

\section{Experiment}

To illustrate the applicability of the method, experimental X-ray speckle tracking data were collected at ESRF beamline BM05, using a red currant sample. The setup corresponds to  Fig.~\ref{fig:ExperimentalSchematic}. The sample was placed on a dedicated stage located 55~m from the source where hard X-ray photons were produced by synchrotron radiation from a 0.85 T dipole bending the trajectory of the 6.02 GeV electrons circulating through the storage ring. The X-ray photon spectral bandwidth was further narrowed to $\Delta E/E \approx 10^{-4}$ and centred around energy $E = 17$~keV using a double crystal Si(111) monochromator located 27 m from the X-ray source. A piece of sandpaper with grit size P800 was fixed on piezo translation motors 0.5~m upstream of the sample.  An imaging detector was placed at a distance $\Delta = 1$~m downstream. This detector consisted of a FReLoN (Fast Read-Out Low-Noise) e2V camera coupled to an optic imaging a thin scintillator \cite{labiche1996frelon,douissard2012}. The effective pixel size of the optical system was 5.8~$\mu$m while also providing a high signal to noise ratio ($> 500$).

% I removed the 'two'
The reference-speckle images were collected, in the absence of the sample, by transversely moving the piece of sandpaper to defined positions of the speckle generator translation motors. Later the images with the sample inserted into the beam were acquired while replacing the sandpaper at precisely the same transverse locations, thanks to the piezo technology of the motors. The images were then processed by running a Python3 code on a simple desktop machine.

%\begin{figure}
%\includegraphics[trim = 62mm 122mm 3mm 18mm, clip, width=0.80\textwidth]{Capture.pdf}
%\caption { (a) Phase Laplacian  $\nabla_{\perp}^2\phi(x,y) $; (b) Recovered dark-field signal $ D_{\textrm{eff}}(x,y;\Delta)$.}
%\label{fig:ExperimentalReconstruction}
%\end{figure}
%%%%%%%%%%%%

\begin{figure*}
\includegraphics[trim =15mm 0mm 15mm 0mm, clip, width=1.0\textwidth]{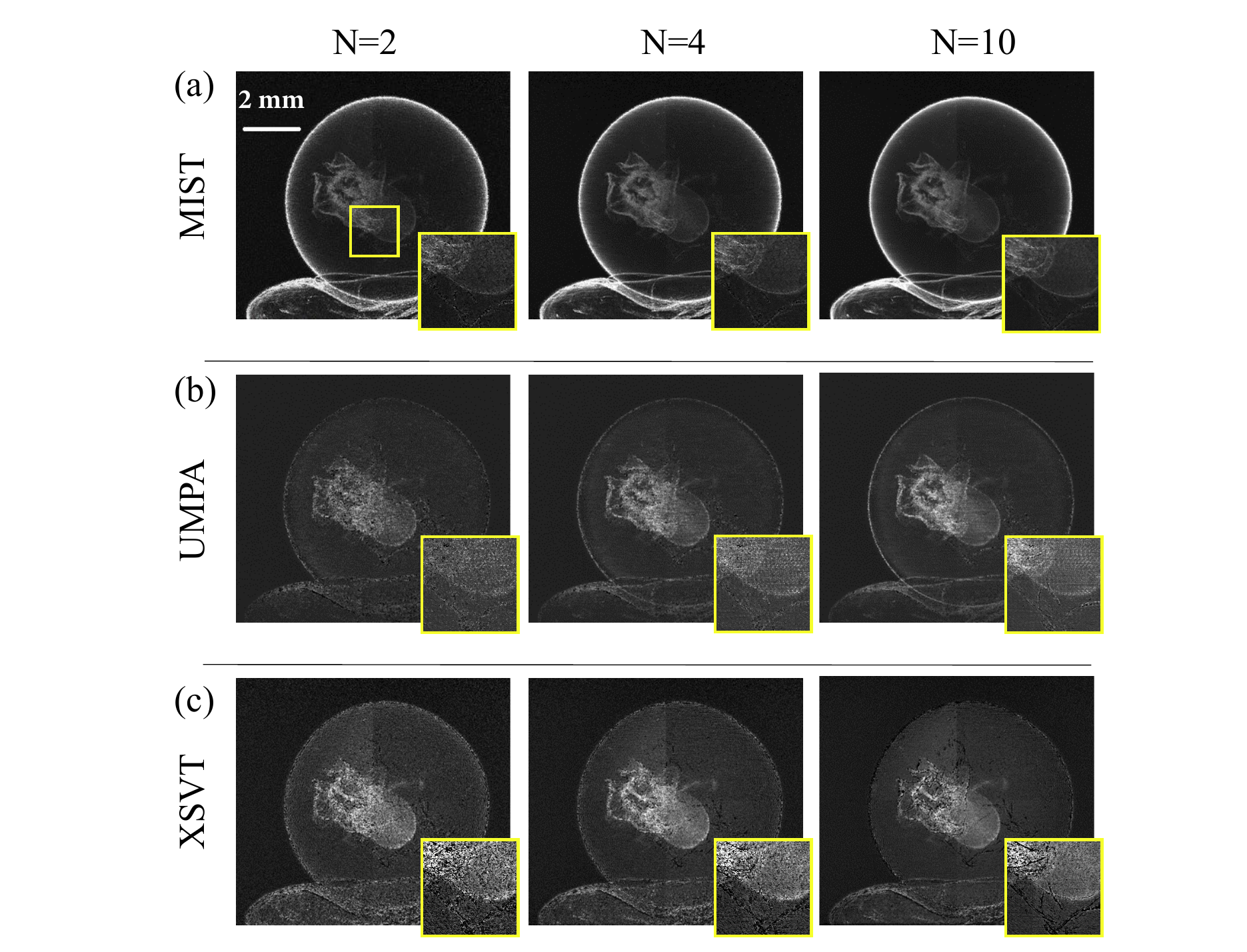}
\caption {(a-c) Recovered dark-field signal $ D_{\textrm{eff}}(x,y;\Delta)$ obtained using the (a) MIST, (b) UMPA and (c) XSVT techniques, with $N=\{2,4,10\}$ pairs of images. The chosen window size in the UMPA and XSVT techniques was equal to $5 \times 5$ pixels.}
\label{fig:ExperimentalReconstruction}
\end{figure*}
% KMP changed Fig.2 by removing the phase info
% (a) Phase Laplacian  $\nabla_{\perp}^2\phi(x,y)$ and (b) phase shift $\phi(x,y)$ in radians. Subsequently a part describing that part of Fig2 was removed.
% Figure \ref{fig:ExperimentalReconstruction}(a) shows the phase Laplacian $\nabla_{\perp}^2\phi(x,y)$ that is given by the upper line of Eq.~(\ref{eq:5}) and in (b) the recovered phase obtained by integrating the result using a fast Fourier transform approach (see e.g.~\citeauthor{GureyevNugent1997} \cite{GureyevNugent1997}) to yield $\phi(x,y)$ up an arbitrary additive constant.

Figure~\ref{fig:ExperimentalReconstruction}(a) shows the positive part of the MIST dark-field signal, $D_{\textrm{eff}}(x,y;\Delta)$, obtained using  Eq.~(\ref{eq:5}) for the case where $N=2$ pairs of images are used. To improve the dark-field signal quality and robustness to noise with the present technique and to compare it to UMPA and XSVT which require a larger number of speckle image pairs, we also calculated similar images using $N=4$ and $N=10$ pairs of speckle images. To do so, we built a system of equations similar to those in Eq.~(\ref{eq:4}), but having $N$ equations corresponding to $N$ different transverse positions of the speckle-generating mask, rather than only two equations (corresponding to two different transverse positions of the speckle-generating mask). We then solved the system of $N$ equations for $\nabla_{\perp}^2\phi(x,y)$ and $D_{\textrm{eff}}(x,y;\Delta)$, by matrix inversion for each pixel, i.e. in the least-squares sense \cite{Press1996} accounting for noise. 

For comparison with the MIST dark-field reconstructions that are presented in Fig.~\ref{fig:ExperimentalReconstruction}(a), Figs.~\ref{fig:ExperimentalReconstruction}(b,c) show corresponding dark-field images obtained using the UMPA and XSVT methods, respectively.   
%To assess the images quality, we calculated both Contrast to Noise Ratios and NIQE image quality metrics \cite{Mittal2013}. 
%Please note that the higher the Contrast to Noise Ratio (CNR) is, the better the image quality is.% whilst it is the opposite with the NIQE metric.
To facilitate further comparison, for dark-field reconstructions obtained using all three methods (MIST, UMPA, XSVT) the Contrast to Noise Ratio (CNR) has been calculated for each reconstruction, as: 
\begin{equation}
\textrm{CNR}=\frac{\mu _1 - \mu _2}{\sigma _1}.
\end{equation}
Here, $\mu _1$ and $\mu _2$ are average intensities measured in an homogeneous zone outside the sample and in the inset of Fig.~\ref{fig:ExperimentalReconstruction} respectively. The parameter $\sigma _1$ refers to the standard deviation measured in the homogeneous zone. Note that higher values of CNR indicate better image quality.  
%For any number of pair image, MIST already visually surpasses the two other methods as the two different metrics indicate. % KMP: 24_08_20: I rewrote this sentence.
For a small number of image pairs ($N=2$ and $N=4$), MIST surpasses the two other methods as indicated by the CNR metric given in Table~\ref{tab:CNR}. 

In the interior of the sample there is broad agreement in the dark-field features that are evident in the reconstructions obtained using the three different methods.  This consistency in the three rows of Fig.~\ref{fig:ExperimentalReconstruction} is notable, since the three corresponding reconstruction formalisms (MIST, UMPA, XSVT)---while of course all ultimately based on the same underpinning optical physics---are conceptually and numerically rather different approaches.  This consistency is also notable on account of the fact that the domain of applicability, of the UMPA and XSVT formalisms, is broader than the domain of validity of MIST.  To counterpoint these consistencies, one can notice that the main difference between MIST and UMPA/XSVT dark-field reconstructions is the more highly contrasted edges of the features with the former technique. % comments on the edge contribution into darkfield  
% added text, KMP 24_08_20

%To assess the images' quality, we calculated the Structural Similarity Index (SSIM) for each of the images obtained with the three techniques. The MIST image obtained from 10 projections was used as reference. The SSIM values are reported below each of the darkfield images in Fig. \ref{fig:ExperimentalReconstruction}.  For any number of pair image, MIST already visually surpasses the two other methods.  One can notice the main difference between MIST and UMPA/XSVT are the more contrasted edges of the features with the former technique, especially for the XSVT technique that almost doesn't account for the contribution providing lower SSIM values. % comments on the edge contribution into darkfield  

\begin{table}[]
    \centering
    \begin{tabular}{c|c|c |c}
         \textbf{CNR} & \textbf{MIST}	& \textbf{UMPA}	& \textbf{XSVT} \\
         \hline
         $N$=2& 10.9	&4.6&	6.9 \\
         $N$=4& 22.6	& 17.9& 	11.6 \\
         $N$=10&23.1&30.8&17.1 \\	
    \end{tabular}
    \caption{Contrast to Noise Ratio comparison for the different dark-field extraction methods}
    \label{tab:CNR}
\end{table}

%\begin{table}[]
%    \centering
%    \begin{tabular}{c|c|c |c}
%         \textbf{Niqe} & \textbf{MIST}	& \textbf{UMPA}	& %\textbf{XSVT} \\
%         \hline
%         n=2& 10.9 &	11.2& 	12.2 \\
%         n=4& 8.8 &	8.9&	10.5 \\
%         n=10 & 8.5&	8.3&	9.6 \\	
%    \end{tabular}
%    \caption{Niqe quality metrics comparison for the %different darkfield extraction}
%    \label{tab:NIQE}
%\end{table}

\section{Discussion}

These obtained results are, to our knowledge, the first experimental implementation of the multi-modal X-ray Fokker--Planck speckle-tracking approach due to \citeauthor{PaganinMorgan2019} \cite{PaganinMorgan2019}. This variant of multi-modal speckle-based X-ray imaging reconstruction takes only a few seconds, which is significantly faster than the XSVT and UMPA approaches. Nevertheless, the results shown in Fig.~\ref{fig:ExperimentalReconstruction}(a) appear to outperform the results obtained from the same experimental data using more sophisticated approaches (see e.g., Fig.~\ref{fig:ExperimentalReconstruction}(b,c) and Fig.~7 in the paper by \citeauthor{Berujon2015c} \cite{Berujon2015c}). % KMP added "outperformed" in the text above, because a word was missing (24_08_20)

Taking into account that the MIST method, described in the present paper, is based on several strong assumptions, the obtained results may contain some artefacts. However, the results obtained by this fast deterministic approach can be used as a starting point for further refinement using more sophisticated (and general) correlation-based techniques, such as XSVT and UMPA.  There is an evident trade-off here: (i) XSVT and UMPA have the advantage of greater generality, which comes at the cost of requiring additional images and significantly longer computation times, while (ii) the MIST method of the present paper has the advantage of requiring fewer images and having much more rapid computation times, at the cost of a reduced degree of  generality.      

It is also worth pointing out that, as our new technique does not explicitly determine the transverse shift of the speckle grains due to phase contrast, we are not restricted by the grain size. The only requirement is that the chosen experimental parameters, namely, the energy, propagation distance and grain size of the mask, should produce a speckle (reference) image.  Rather than the resolution of the MIST method being determined by the size of the speckles, it is determined by the average  characteristic length scale that is present in the reference speckle image.  For smooth speckles in which there is only one characteristic transverse length scale, the size of the speckles will be equal to the average characteristic length scale that is present in the reference speckle images.  However, more generally speaking, these length scales may be different, e.g.~if (i) the speckles themselves have sharp edges (as can be the case when the speckles are generated via absorption from suitable powders or grains); or (ii) one has ``speckled speckle'' \cite{Freund2008}.  In general, the speckle size provides an upper limit on this characteristic transverse length scale.

Let us now return to a point raised at the end of the previous section.  Notwithstanding the strong degree of similarity between the dark-field reconstructions in Fig.~\ref{fig:ExperimentalReconstruction}, one point of difference is in the stronger contrast that is observed at the edges of the berry, in the MIST dark-field reconstructions shown in Fig.~\ref{fig:ExperimentalReconstruction}(a), when compared to the UMPA and XSVT reconstructions shown in Figs.~\ref{fig:ExperimentalReconstruction}(b,c).  We believe this difference to be related to the fact that in the Fokker--Planck formalism, upon which the approach of the present paper is predicated, the position-dependent effective diffusion coefficient has three separate contributions.  The first contribution to the effective diffusion coefficient is the local SAXS induced by spatially-unresolved sample micro-structure, with this contribution being treated in an ultimately-equivalent albeit conceptually-different manner in MIST, UMPA and XSVT.  The second contribution to the Fokker--Planck effective diffusion coefficient \cite{MorganPaganin2019,PaganinMorgan2019,GureyevPMB2020} is the Young--Maggi--Rubinowicz boundary wave \cite{YoungOnTheBoundaryWave,Maggi,Rubinowicz,MiyamotoWolf1,MiyamotoWolf2} associated with diffraction from sharp intensity gradients such as sample edges \footnote{Cf.~the use of such an edge wave in the different phase-retrieval context of a deterministic approach to coherent diffractive imaging \cite{Podorov2007,Pavlov2017,Pavlov2018}. This may also be analogous to the different type of behaviour observed for diffuse scattering from dislocation ensembles having shorter and longer correlation lengths. The former (weak correlations, e.g., for a random distribution of deformation sources) produce a wider distribution of intensity in reciprocal space, while the latter (strong correlations, e.g., dislocation walls/nets) have narrower peaks, which is somewhat similar to the boundary effect mentioned in the main text of the paper.  See e.g.~Krivoglaz \cite{Krivoglaz1,Krivoglaz2} as well as Kaganer and Sabelfeld \cite{Kaganer}, together with references therein, for further information regarding this connection.}.  This second contribution may also be viewed in ray-optics language as being due to Keller-type ``diffracted rays'', under the viewpoint of the geometrical theory of diffraction \cite{Keller}.  Yet another way of modelling this second contribution is via the asymptotic form of two-dimensional diffraction integrals, in which sharp scattering edges constitute a critical point of the second kind \cite{MandelWolf}.  For a clear example of the diffusive effect of edge-scattered rays, see e.g.~Fig.~2(b) in \citet{Groenendijk2020}, where the illuminating grid pattern has reduced visibility in the vicinity of a sharp sample edge \footnote{The observation, made in this sentence, was pointed out to the authors in a private communication from K.~S.~Morgan to DMP, on September 9, 2020.}. Regardless of the adopted physical model, the key point regarding this second (edge-diffraction-based) contribution to the effective position-dependent Fokker--Planck diffusion coefficient, is that it provides a distinct mechanism for diffusive photon transport, which is different from the local-SAXS mechanism.  While the local-SAXS mechanism is ultimately equivalently treated in the dark-field signal reconstructed by MIST, UMPA and XSVT, the methods treat the edge signal differently, an observation that is consistent with the fact that (i) all three methods give very similar contrast for the dark-field signal that is reconstructed within the interior of the berry, in all panels of Fig.~\ref{fig:ExperimentalReconstruction}, but (ii) MIST gives a reconstruction that differs from UMPA and XSVT, at the edges of the berry.  Finally, we mention a third contribution to the effective diffusion coefficient in the Fokker--Planck formalism, namely that which is due to the effective point-spread function associated with the imaging system \cite{MorganPaganin2019,PaganinMorgan2019}.  This third contribution is listed for completeness, but is otherwise of marginal relevance here, since its contribution will be negligible in the present context of a Fokker--Planck optical-flow approach to X-ray speckle tracking.  

It would be interesting to extend the Fokker--Planck speckle-tracking formalism, upon which the present work is based, to the case of ``directional dark field'' \cite{jensen2010directional} in which the transverse cross section of the local SAXS fan is modelled as being elliptical rather than rotationally symmetric.  In this case, the scalar effective diffusion coefficient may be replaced with a symmetric rank-two diffusion tensor \cite{MorganPaganin2019,PaganinMorgan2019}:
\begin{eqnarray}
D_{\textrm{eff}}(x,y;\Delta)\longrightarrow \begin{bmatrix}
D^{(xx)}_{\textrm{eff}}(x,y;\Delta) & 
D^{(xy)}_{\textrm{eff}}(x,y;\Delta) \\
D^{(xy)}_{\textrm{eff}}(x,y;\Delta) & D^{(yy)}_{\textrm{eff}}(x,y;\Delta)
\end{bmatrix} . ~
\label{eq:DiffusionTensor}
\end{eqnarray}
Following Eq.~(4) in \citet{MorganPaganin2019} and Eq.~(51) in \citet{PaganinMorgan2019}, the above diffusion tensor enables us to write down the following directional-dark-field generalisation of Eq.~(\ref{eq:1}):  %
\begin{eqnarray}
  \nonumber I_{\textrm R}(x,y) - I_{\textrm S}(x,y) &=& \frac{\Delta}{k} \nabla_{\perp}\cdot [I_{\textrm R}(x,y) \nabla_{\perp}\phi(x,y)] \\ \nonumber &-& \Delta\frac{\partial^2}{\partial x^2}[D^{(xx)}_{\textrm{eff}}(x,y;\Delta)I_{\textrm R}(x,y)]  
  \\ \nonumber &-& \Delta\frac{\partial^2}{\partial y^2}[D^{(yy)}_{\textrm{eff}}(x,y;\Delta)I_{\textrm R}(x,y)]  
  \\ \nonumber &-& \Delta\frac{\partial^2}{\partial x\partial y}[D^{(xy)}_{\textrm{eff}}(x,y;\Delta)I_{\textrm R}(x,y)]. \\
  \label{eq:1--generalised-form} 
\end{eqnarray}
The above equation could be readily solved, in an analogous manner to the method given in the present paper, thereby extending MIST into a method capable of measuring a directional dark-field signal, as quantified by the diffusion tensor in Eq.~(\ref{eq:DiffusionTensor}).  Such an investigation would be an interesting avenue for future work.

\section{Conclusion}

We have developed a fast deterministic variant of X-ray Multi-modal Intrinsic-Speckle-Tracking (MIST), which was validated using experimental data. The obtained reconstruction results %for the object's refractive and SAXS properties 
are based on only two images of the sample acquired at different positions of the spatially random mask. These reconstructions are comparable in quality to those obtained by computationally slower (multiple-image) albeit significantly more general, explicit X-ray speckle tracking techniques.

\section*{Acknowledgements}
We acknowledge the European Synchrotron Radiation Facility for provision of synchrotron radiation facilities. We acknowledge useful discussions with Scott Findlay, Andrew Kingston, Marcus Kitchen, Thomas Leatham, Kaye Morgan, Gary Ruben, Sameera Tadikonda and Marie-Christine Zdora.

\bibliography{MIST}% Produces the bibliography via BibTeX.

\end{document}